\documentclass[runningheads]{llncs}
\usepackage{graphicx}
\usepackage{cite}
\usepackage{url}
\usepackage{marvosym}

\usepackage{subcaption}                      %
\begin{document}
\title{A SARS-CoV-2 Microscopic Image Dataset with Ground Truth Images and Visual Features}
%
%
\author{Chen Li\inst{1 \textsuperscript{\Letter}}\and
Jiawei Zhang\inst{1}\and
Frank Kulwa\inst{1}\and
Shouliang Qi\inst{1}\and
Ziyu Qi\inst{2}}

\authorrunning{C. Li et al.}
\titlerunning{SARS-CoV-2 Dataset with GT and Visual Features}
%
\institute{Microscopic Image and Medical Image Analysis Group, MBIE College, 
Northeastern University, Shenyang, 110169, China \\
\email{lichen201096@hotmail.com},
\email{1971087@stu.neu.edu.cn},
\email{frank.kulwa@gmail.com},
\email{qisl@bmie.neu.edu.cn}\and 
School of Life Science and Technology,\\ 
University of Electronic Science and Technology of China, Chengdu, 611731, China \\
\email{qi.ziyu@outlook.com}}
\maketitle              
\begin{abstract}
SARS-CoV-2 has characteristics of wide contagion and quick propagation velocity. 
To analyse the visual information of it, we build a SARS-CoV-2 Microscopic Image Dataset (SC2-MID) with 48 electron microscopic images and also prepare their ground truth images. 
Furthermore, we extract multiple classical features and novel deep learning features 
to describe the visual information of SARS-CoV-2. Finally, it is proved that the 
visual features of the SARS-CoV-2 images which are observed under the electron 
microscopic can be extracted and analysed.

\end{abstract}
\footnotetext[1]{J. Zhang—Cofirst author. This work is supported by ``National Natural Science Foundation of China'' (No. 61806047), the ``Fundamental Research Funds for the Central Universities'' (Nos. N2019003 and N2019005), and the China Scholarship Council (No. 2017GXZ026396). }

\footnotetext[2]{\copyright  Springer Nature Switzerland AG 2020 \\Y. Peng et al. (Eds.): PRCV 2020, LNCS 12305, pp. 244–255, 2020. \\https://doi.org/10.1007/978-3-030-60633-6\_20}

\section{Introduction}

It is reported that the severe acute respiratory syndrome 
coronavirus 2 (SARS-CoV-2) breaks out since the end of December 
in 2019~\cite{hui2020the}. More than 12,690,000 people have been 
infected around the world till July 12th, 2020~\cite{JHU-2020-CCGC}, 
which becomes a global malignant epidemic.

A novel coronavirus was detected for the first time in the 
laboratory on January 7th, 2020, and the whole genome sequence 
of the virus was obtained. The first 15 positive cases of novel 
coronavirus were detected by nucleic acid detection and the 
virus was separated from one positive patient and observed under 
an electron microscope. The detection of pathogenic nucleic acid 
was completed on January 10th, 2020. The first novel coronavirus 
with electron microscopic images in China was successfully 
separated from the Center for Disease Control and Prevention (CDC) 
on January 24th, 2020. Novel coronavirus nucleic acids were 
detected in 33 samples from China's CDC  on January 26th, 2020, 
and the virus was successfully separated from positive environmental 
samples. SARS-CoV-2 outbreak was announced as a public health emergency of international concern (PHEIC) by the World Health 
Organization (WHO) on January 30th, 2020.  It was officially named 
as COVD-19 (corona virus disease 2019) by WHO on Febuary 11th, 2020. 
The International Committee on Taxonomy of Viruses (ICTV) has named 
the virus as Severe Acute Respiratory Syndrome Coronavirus 2 
(SARS-CoV-2)~\cite{gorbalenya2020severe}.

Coronavirus is a kind of plus-strand RNA virus, which can infect many 
mammals including human beings and can cause cold and some other serious 
diseases such as Middle East Respiratory Syndrome (MERS) and Severe Acute 
Respiratory Syndrome (SARS)~\cite{cui2019oringin}. The SARS-CoV-2 is a 
novel coronavirus that has never been found in human bodies. The SARS-CoV-2 
is highly infectious and mainly transmitted through close contact and 
respiratory droplets. Besides, the severe patients may 
die~\cite{malik2020emerging}. The main symptom of human infection is 
respiratory disease, accompanied by fever, cough and may cause viral 
pneumonia\cite{chen2020epidemiological}. There is no effective medicine 
developed up to now. Because of the rapid mutation of RNA coronavirus 
and lack of efficient medicine, there are mainly two methods to confirm 
the case, the first is detecting the positive nucleic acid of coronavirus 
by using RT-PCR, the other one is the high homology between the results 
of the virus gene sequence and the known SARS-CoV-2~\cite{chu2020molecular}. 
Both of the methods above need professional medical equipments and personnel. 
So the process is time consuming and expensive.  Microscopic image analysis 
provides a new method for rapid coronavirus screening~\cite{Li-2019-ASS,Li-2019-ASF,Li-2016-CBMIA,Li-2020-AROC,Li-2018-ABR}, thus, several visual features are extracted 
in the experiments below.

\section{ SARS-CoV-2 Microscopic Image Dataset (SC2-MID)}

\subsection{ SARS-CoV-2 Visual Properties and Ground Truth Image Preparation}

\begin{figure}[h]
\centering
\includegraphics[trim={0cm 4cm 0cm 9cm},clip,width=\textwidth]{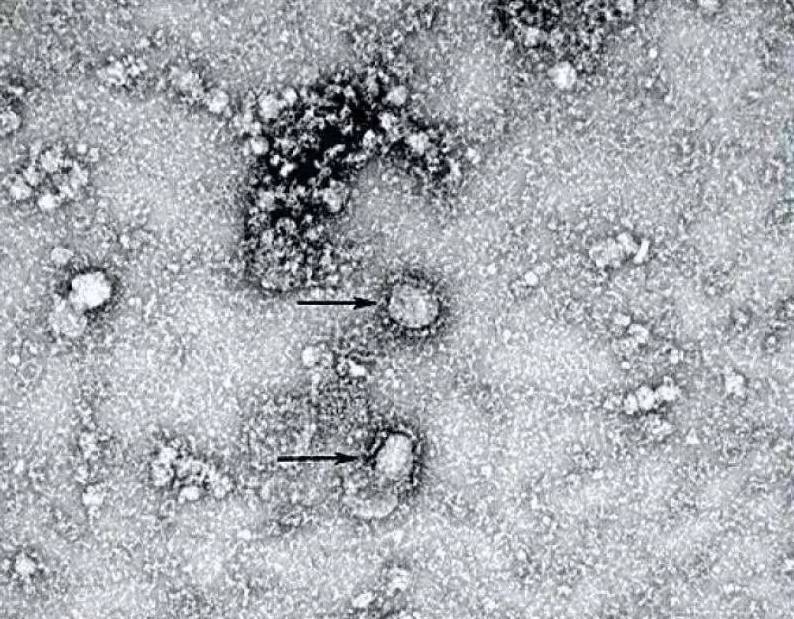}
\caption{The electron microscopic image of SARS-CoV-2\cite{ baidu2020the}.}
\label{fig:2019-nCOV01}
\end{figure}

SARS-CoV-2 is a type of $\beta$ coronavirus, which has envelope with 
spinous process. The shape of the virus particle is circular or oval, 
which seems as a solar corona. The tubular inclusions can be detected 
in the cell which is infected with coronavirus. The spinous 
process of different coronaviruses has significant differences. The 
SARS-CoV-2 has a diameter of 60-140nm. It also has the largest genome as 
a RNA virus. The mircoscopic image of SARS-CoV-2 is shown in 
Fig.~\ref{fig:2019-nCOV01}. 

\begin{figure}[h]
 \centering
\begin{subfigure}[t]{0.45\textwidth}
  \includegraphics[width=1\textwidth]{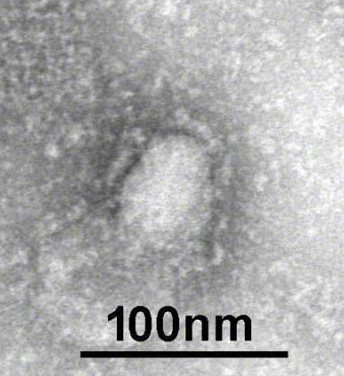}
 \caption{Original Image.}
 \end{subfigure}%
 \hspace{0.2cm}
 \begin{subfigure}[t]{0.45\textwidth}
  \includegraphics[width=1\textwidth]{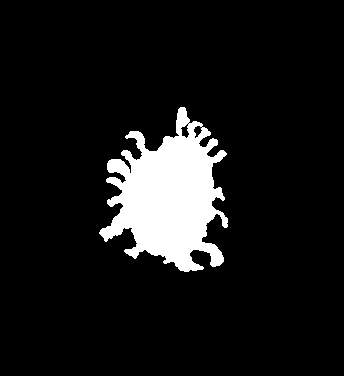}
 \caption{Ground Truth Image.}
 \end{subfigure}
\caption{An example of original image and ground truth image.}
\label{fig:originalgt}
\end{figure}

In order to obtain the accurate image and extract the visual features of SARS-CoV-2, a dataset is constructed and one of the example images with its Ground Truth (GT) image are shown in Fig.~\ref{fig:originalgt}. Geometric features and texture features of SARS-CoV-2 are extracted by combining both the original image and GT image.

Based on the visual properties of SARS-CoV-2 images, we prepare the GT 
images according to the rules as follows:

\subsubsection{a}
The GT images in the dataset are generated in pixel level. The foreground of GT 
images is the SARS-CoV-2 and the pixel value is ``1" which represents white. The pixel value of the background is zero (``0") which represents black.

\subsubsection{b}
Then all part of SARS-CoV-2 is drawn from the observation by naked eyes.

\subsubsection{c}
When the spinous region of SARS-CoV-2 is not clear, the GT image 
is drawn following the value of the particle region. If the SARS-CoV-2 particle is 
bright part, then the bright region around the particle will be drawn as spinous part of SARS-CoV-2, and vice versa.

\subsection{Dataset Construction}

Most of the SARS-CoV-2 images are kept by National Health Commission 
at present. One of the images with scale is shown in Fig.~\ref{fig:originalgt}(a)
and the corresponding GT image is shown in Fig.~\ref{fig:originalgt}(b). 
The scale of Fig.~\ref{fig:originalgt}(a) is labelled in the image precisely 
and can be used to calculate the real size of the SARS-CoV-2.

The SARS-CoV-2 microscopic image dataset (SC2-MID) we built has 17 electron microscopic images which are separated to 48 single SARS-CoV-2 images. We are willing to share the dataset with the researchers for sharing but not for commercial purposes. If you want to obtain the dataset, please contact our data manager Jiawei Zhang (1971087@stu.neu.edu.cn). The horizontal axes of images are roughly the same length with the scale length in Fig.~\ref{fig:originalgt}(a) 
which has a length of 100$nm$. The other images are resized and cut based on the scale of Fig.~\ref{fig:originalgt}(a). The image names with their data sources are shown in 
Table~\ref{tab:source}.

\begin{table}[h] 
\centering
\caption{\label{tab:source}The sources of SARS-CoV-2 dataset.} 
\begin{tabular}{p{5cm}p{7cm}} 
\hline
ImageName & Source \\
\hline
{\scriptsize IMG-001 - IMG-004} & {\scriptsize https://www.infectiousdiseaseadvisor.com/home/topics/gi-illness/covid-19-symptoms-may-need-to-be-extended-to-include-gi-symptoms/ }\\ 
{\scriptsize IMG-005 - IMG-008} & {\scriptsize https://wired.jp/2020/03/08/what-is-a-coronavirus/}\\ 
{\scriptsize IMG-009 - IMG-014} & {\scriptsize https://www.genengnews.com/news/sars-cov-2-insists-on-making-a-name-for-itself/}\\ 
{\scriptsize IMG-015 - IMG-018} & {\scriptsize https://www.h-brs.de/en/information-on-coronavirus}\\ 
{\scriptsize IMG-019 - IMG-022} & {\scriptsize https://www.medicalnewstoday.com/articles/why-does-sars-cov-2-spread-so-easily}\\ 
{\scriptsize IMG-023 - IMG-028} & {\scriptsize https://www.upwr.edu.pl/news/51004/coronavirus-sars-cov-2-messages.html}\\ 
{\scriptsize IMG-029 - IMG-032} & {\scriptsize https://www.infectiousdiseaseadvisor.com/home/topics/gi-illness/covid-19-symptoms-may-need-to-be-extended-to-include-gi-symptoms/}\\ 
{\scriptsize IMG-033 - IMG-035} & {\scriptsize https://www.dzif.de/en/sars-cov-2-dzif-scientists-and-development-vaccines}\\ 
{\scriptsize IMG-036 - IMG-038} & {\scriptsize https://www.charite.de/en/clinical-center/themes-hospital/faqs-on-sars-cov-2/:}\\ 
{\scriptsize IMG-039 - IMG-041} &{\scriptsize https://news.harvard.edu/gazette/story/2020/03/in-creating-a-coronavirus-vaccine-researchers-prepare-for-future/}\\ 
{\scriptsize IMG-042} & {\scriptsize https://newsbash.ru/society/health/17116-kak-vygljadit-koronavirus-pod-mikroskopom-rossijskie-uchenye-sdelali-foto.html}\\ 
{\scriptsize IMG-043} & {\scriptsize https://www.rbc.ru/rbcfreenews/5e735ff09a7947be392f2bec}\\ 
{\scriptsize IMG-044} & {\scriptsize http://www.ellegirl.ru/articles/foto-dnya-kak-vyglyadit-koronavirus/}\\ 
{\scriptsize IMG-045 - IMG-046} & {\scriptsize https://mp.weixin.qq.com/s/zO8rW8W2TgzN2o6JEKcLnQ}\\ 
{\scriptsize IMG-047 - IMG-048} & {\scriptsize https://br.sputniknews.com/asia-oceania/2020012415043397-china-publica-foto-do-coronavirus-visto-por-microscopio-eletronico/}\\ 
\hline
\end{tabular} 
\end{table}

\begin{figure}[h]
 \centering
\begin{subfigure}[t]{1\textwidth}
  \includegraphics[width=1\textwidth]{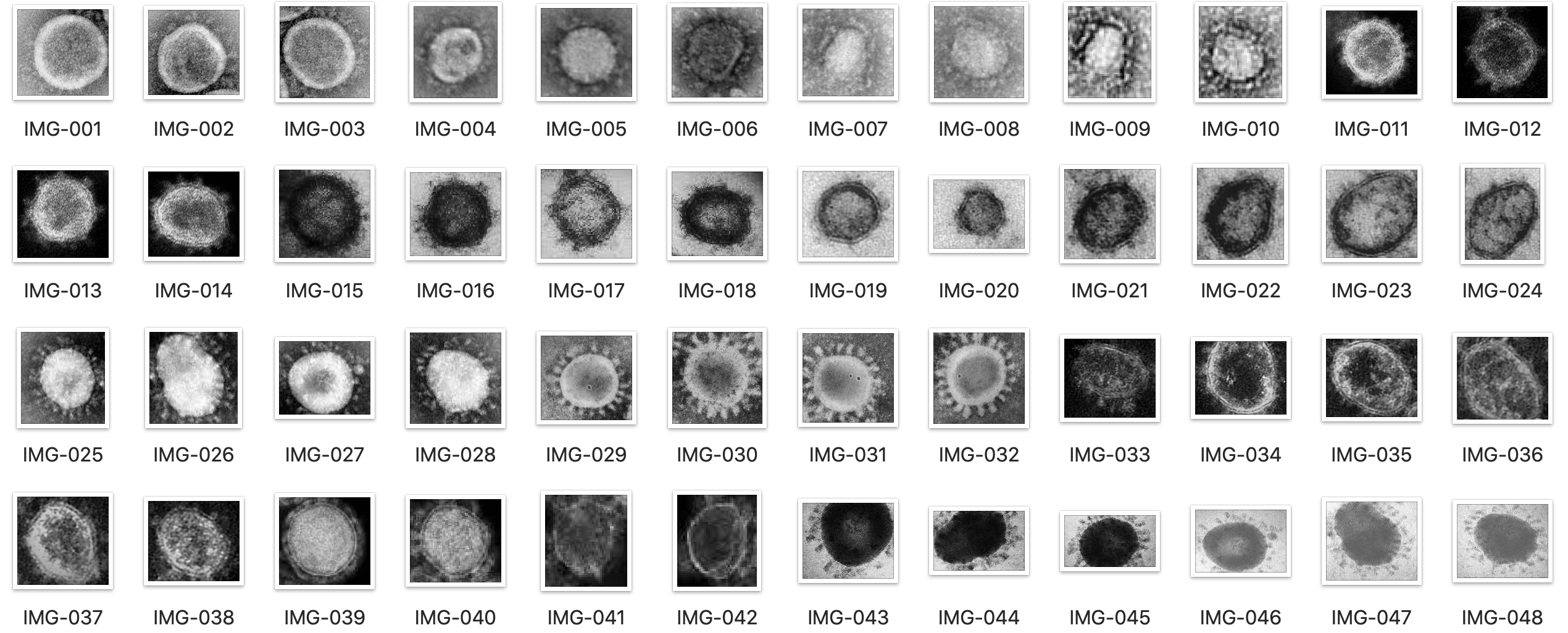}
 \caption{Original images.}
 \label{fig:results55a}
 \end{subfigure}%

 \begin{subfigure}[t]{1\textwidth}
  \includegraphics[width=1\textwidth]{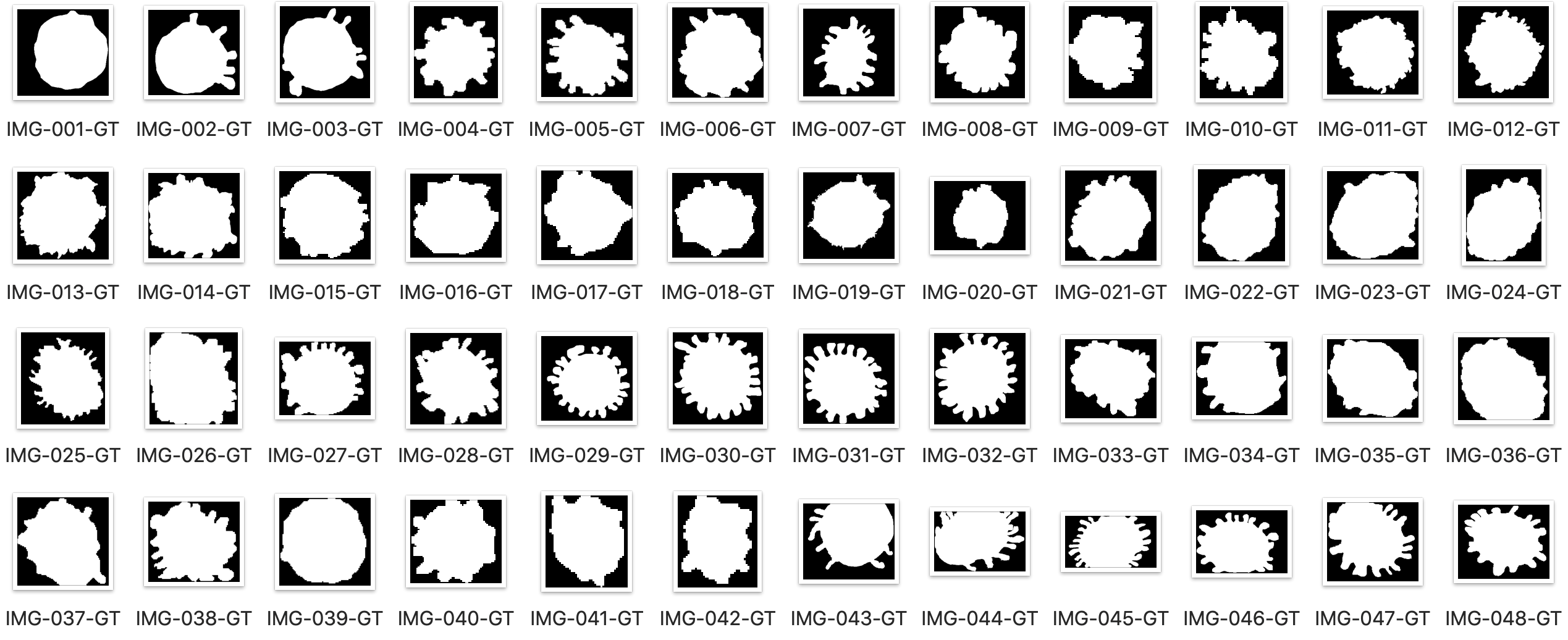}
 \caption{Ground truth images.}
 \end{subfigure}
\caption{The database of gray images and GT images of SARS-CoV-2.}
\label{fig:gray-gt}
\end{figure}

The database of SARS-CoV-2 is built which contains 48 electron microscopic images. The corresponding GT images are also shown in Fig.~\ref{fig:gray-gt}.

\section{SARS-CoV-2 Visual Feature Extraction}

Visual feature extraction is one of the important parts of computer vision. 
We extract several shape features including basic geometric feature and Hu 
invariant moment. We also extract several texture features including Histogram 
of Oriented Gradient (HOG) and Gray-Level Co-occurrence Matrix (GLCM). And 
we use deep learning such as VGG-16, Xception and DenseNet121 to get the feature map of images. All the features above can be extracted from the dataset we constructed. An example of feature extraction is shown as follows.

\subsection{Shape Feature Extraction}

Shape feature extraction is one of the most important research topics 
in describing the true nature of images. The original shape 
of an object can be precisely stored by using shape features, which is 
significant for computer vision and image recognition.

\subsubsection{Basic Geometric Features:}

Perimeter refers to the boundary length of the object in the image. 
The perimeter of the image is composed of several discrete pixel points, 
which is calculated as the sum of pixel points of the target edge.

The area of an object in an image is usually represented by calculating the 
sum of pixel points of the object. The GT image is used to calculate the 
area of SARS-CoV-2 in the original image. The SARS-CoV-2 in GT image is shown 
as white which is represented by 1, and the background is shown as black 
which is represented by 0. Scan the GT image and sum the number of pixel 
points that the value is 1. The process is defined as below:
\begin{equation}
   A=\sum_{x=1}^{N}\sum_{y=1}^{M}f(x,y),
\end{equation}
The area of SARS-CoV-2 is represented by the sum of the pixels that $f(x,y)=1$.

The major axis and the minor axis are the longest length and the shortest 
length while linking two random points of an oval. They are usually 
represented as the major axis and minor axis of the smallest oval which 
can contain all of the objects in an irregular figure. The major axis and 
minor axis are defined as below:
\begin{equation}
   l_{a}=max{\sqrt{(i_{m}-i_{n})^{2}+(j_{m}-n_{n})^{2}}},
\end{equation}
where $l_{a}$ is the length of the major axis, $i_{m}$, $i_{n}$, $j_{m}$, $j_{n}$ are boundary points of the connected region in four directions. 
\begin{equation}
   s_{a}=\frac{p}{t}-l_{a},
\end{equation}
where $s_{a}$ is the length of minor axis, $p$ is the perimeter, $t$ is 
the ovality factor defined by $s_{a}/l_{a}$. 

From Fig.~\ref{fig:originalgt}, we segment the image scale and match it 
to pixels. The pixels of part of the scale are black and the rest pixels of 
the image are white. So it is easy to calculate the length of scale and 
calculate the proportion of the image size to true size. And then the 
proportion can be used to calculate the values of the shape features. 
The length of scale consists of 206 pixels which means 100nm for real size, 
so the proportion is about 0.4854 nm/pixels. 

\begin{table}[h] 
\centering
\caption{\label{tab:shape}The shape feature values of SARS-CoV-2.} 
\begin{tabular}{p{5cm}p{5cm}p{2cm}} 
\hline
Feature & Value & Unit \\
\hline
Area & 2808.98 & $\rm nm^{2}$\\ 
Perimeter & 477.4689 &  $\rm nm$ \\ 
MajorAxis & 71.5500 &  $\rm nm$ \\
MinorAxis & 56.8100 &  $\rm nm$ \\
Eccentricity & 1.2595 & --\\
\hline
\end{tabular} 
\end{table}

The shape features of SARS-CoV-2 are extracted by combining the GT image 
and original image, and the true size of these features are calculated by 
using the proportion above and the values are shown in Table~\ref{tab:shape}. 
The eccentricity is the ratio of the major axis to the minor axis.

\subsubsection{ Hu Invariant Moment:}

Hu invariant moment creates seven invariant moment functions by using the normalized 
second and third order center distance. The central moment is defined as bellow:
\begin{equation}
  \mu_{pq}=\sum_{x=1}^{M}\sum_{y=1}^{N}({x-x_{0})^{p}({y-y_{0})^{q}f(x,y)}},
\end{equation}
where $(x_{0},y_{0})$ is the center of gravity coordinates of the image.

HU is used to describe the properties of images. The Hu invariant moment
is widely used because of its high stability while changing the geometric 
characteristics of the images, which is invariant to translation, rotation and scale transformation. 
Hu invariant moment is generally used to identify large objects in an image, which
can describe the shape of objects well and recognize them quickly. The seven values of Hu 
invariant moment for Fig.~\ref{fig:originalgt} are shown in Table~\ref{tab:hu}. 

\begin{table}[h] 
\centering
\caption{\label{tab:hu}The Hu invariant monent values of SARS-CoV-2.} 
\begin{tabular}{p{1.6cm}p{1.6cm}p{1.6cm}p{1.6cm}p{1.6cm}p{1.6cm}p{1.6cm}} 
\hline
0.9584 & 0.1854 & 0.0270 & 0.0125 & -0.0004 & -0.0021 & -0.0002\\
\hline
\end{tabular} 
\end{table}

The geometric values are roughly consistent with the morphology of the virus 
when observed by the naked eyes. It is much more accurate to obtain the 
geometric values of the virus through the GT image. The geometric values are 
all objective values obtained by computer, but the conclusion is a little 
more subjective. 

\subsection{Texture Feature Extraction}

\subsubsection{HOG Feature Extraction:}
HOG is a classical method to recognize 
the object and extract the texture features\cite{n2005histograms}. The 
local HOG can describe the texture features of particular part in an image. 
The principle of HOG is selecting the gradient of image edge area and 
extracting the density distribution coefficient. The first step is dividing 
the global image into several sub-images based on pixels. Then calculate 
the oriented gradient values and save them into a matrix. Finally, integrate 
the matrix based on initial image.

\begin{figure}[h]
\centering
\includegraphics[trim={0cm 0cm 0cm 0cm},clip,width=0.9\textwidth]{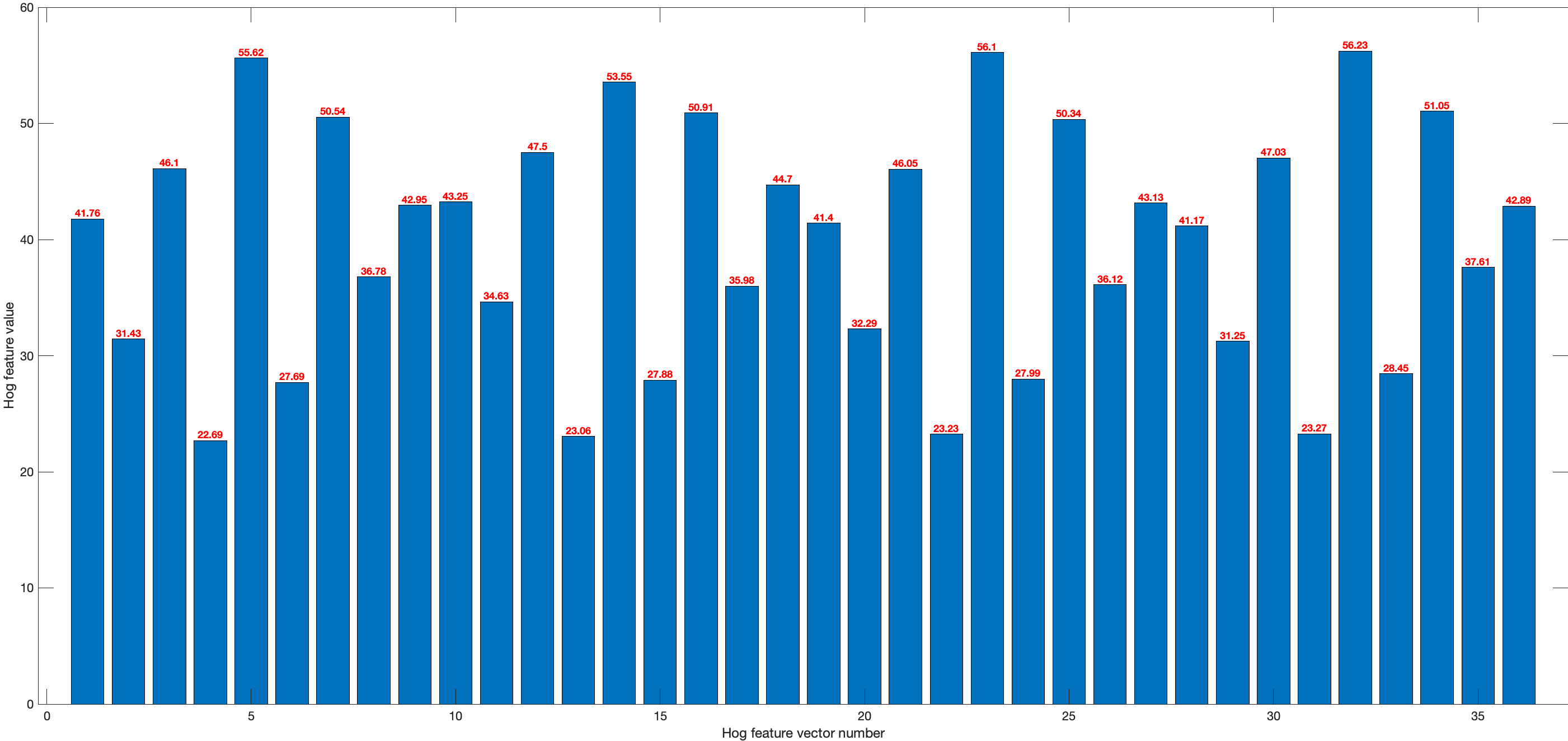}
\caption{HOG histogram of SARS-CoV-2.}
\label{hog_num}
\end{figure}

Because of the rich information and high diversity of image,
the dimensions of extracted HOG feature are different. So it is necessary to normalize 
the HOG feature into a 36 dimensional vector, which has 4 blocks in a region 
with 9 dimensions per block. We extract the HOG feature of SARS-CoV-2 by combining 
the GT and original images. The normalized histogram with 36 dimensional vectors is 
shown in Fig.~\ref{hog_num}.

\begin{figure}[h]
\centering
\includegraphics[trim={0cm 6cm 0cm 6.5cm},clip,width=0.85\textwidth]{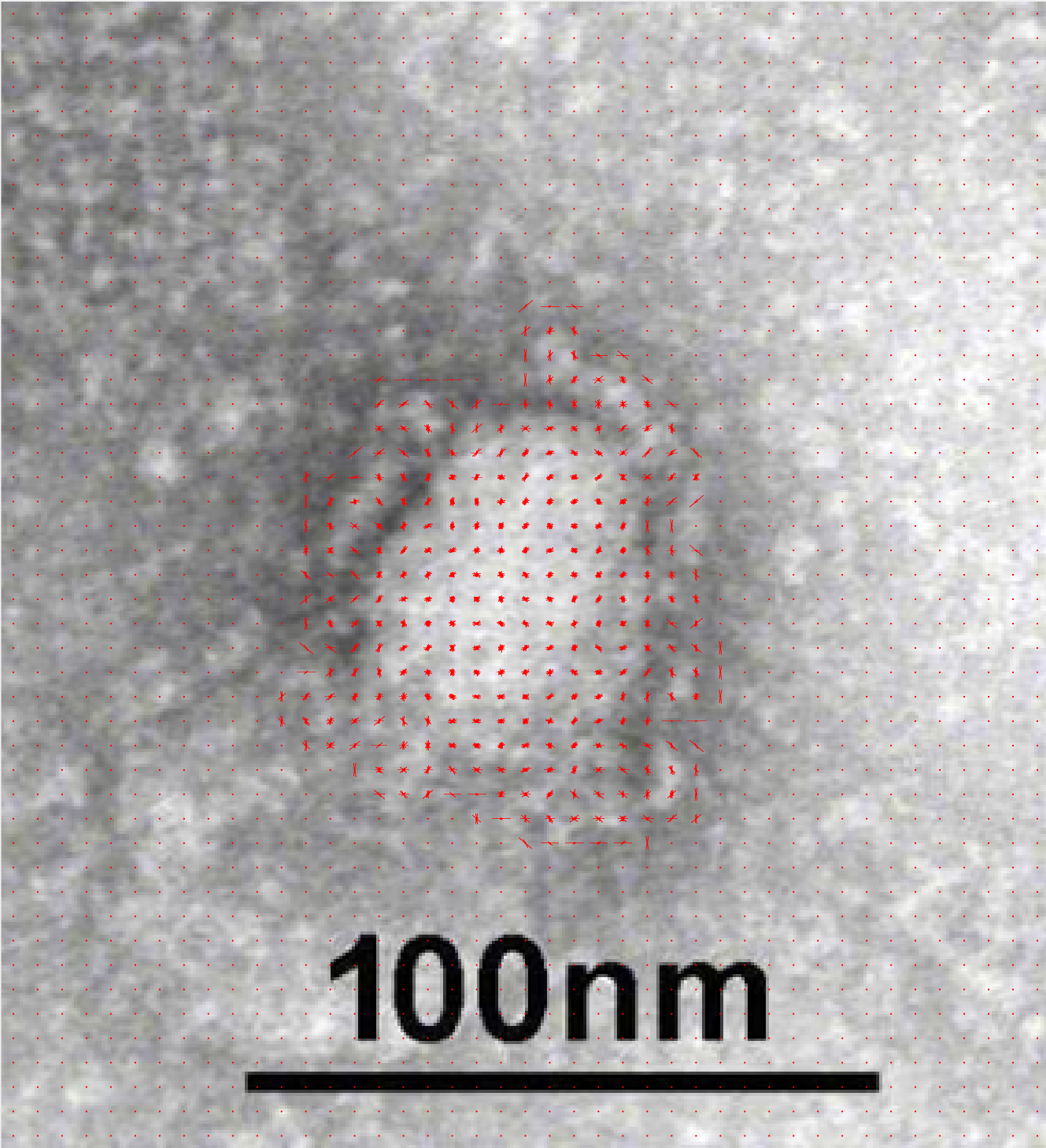}
\caption{The vector graph of HOG feature for SARS-CoV-2.}
\label{virus_hog2}
\end{figure}

The vector graph of HOG feature is shown in Fig.~\ref{virus_hog2}. The 
red arrows describe the change of oriented gradient precisely. The vector 
graph shows good ability in describing the HOG feature of the part of 
SARS-CoV-2.

\subsubsection{GLCM Feature Extraction:}
\begin{figure}[h]
\centering
\includegraphics[trim={0cm 0cm 0cm 0cm},clip,width=0.9\textwidth]{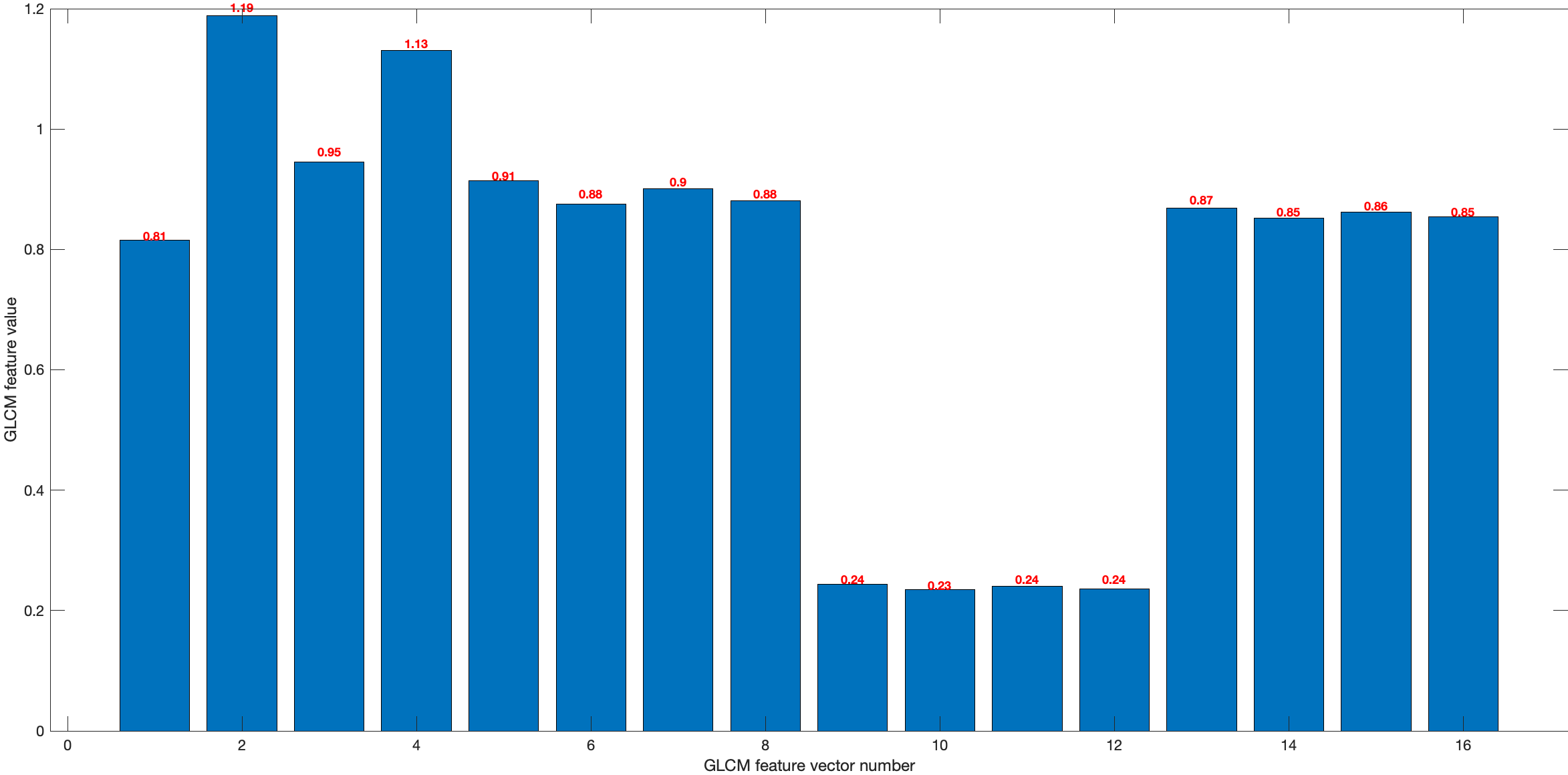}
\caption{GLCM histogram of SARS-CoV-2.}
\label{virus_glcm2}
\end{figure}

GLCM is one of the general methods to describe the texture features of images. The principle is measuring the spatial information between two pixels to describe the texture features. The texture feature refers to the gray-scale relationship between two pixels. Find the corresponding relationship between the pixel and the pixels in eight directions. The GLCM is to combine the co-occurrence matrix between every two pixels in the image. There are four kinds of eigenvalues in GLCM, which are contrast, homogeneity, correlation and energy.Every feature above has 4 dimensional vectors. The normalized histogram with 16 dimensional vectors is shown in Fig.~\ref{virus_glcm2}.

The edge feature of the image can be well 
extracted without losing much local details, and the feature of low sensitivity 
for local geometry and optical transformation can be acquired by HOG and 
GLCM feature extraction, which can express the texture feature of SARS-CoV-2 
effectively.

\subsection{Deep Learning Feature Extraction}
Convolution neural networks (CNN) have the ability to learn deep features which are substituted to hand crafted features such as corners, edges, blobs and ridges. These features are very robust and invariant to image translational changes. 

\begin{figure}[h!]
\centering
\includegraphics[scale=0.8]{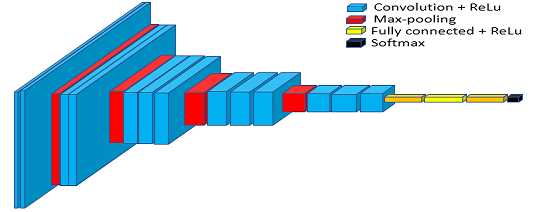}
\caption{VGG-16 network for extracting deep learning features.}
\label{Vgg16}
\end{figure}

\begin{figure}[h]
 \centering
\begin{subfigure}[t]{0.8\textwidth}
  \includegraphics[width=1\textwidth]{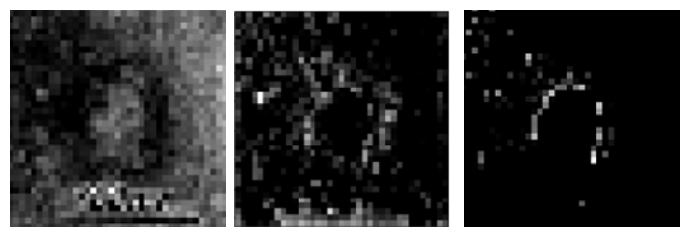}
 \caption{Extraction of edge features of SARS-CoV-2 using Xception CNN.}
 \end{subfigure}%

 \begin{subfigure}[t]{0.79\textwidth}
  \includegraphics[width=1\textwidth]{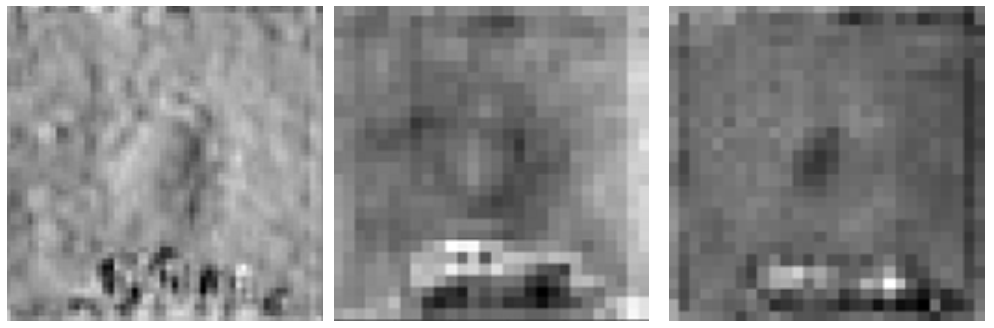}
 \caption{Blurring of the image of SARS-CoV-2 using DenseNet 121.}
 \end{subfigure}

 \begin{subfigure}[t]{0.8\textwidth}
  \includegraphics[width=1\textwidth]{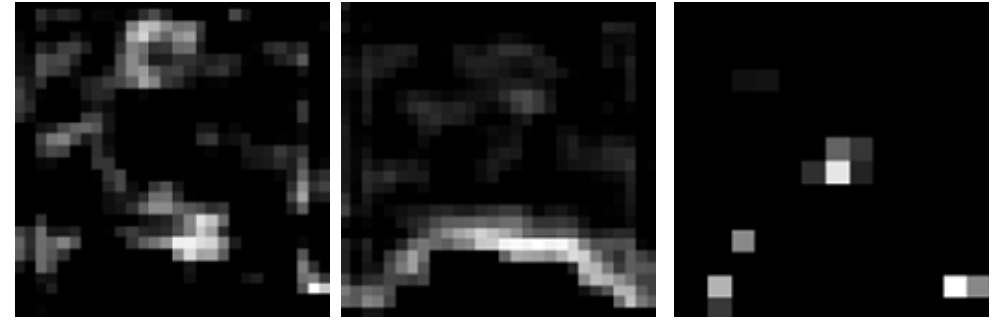}
 \caption{Corner and edge feature learning by VGG-16.}
 \end{subfigure}
\caption{Extracted feature maps of dataset using CNNs.}
\label{fig:deepFeatures}
\end{figure}

CNN uses multiple convolution layers to progressively extract high and low level features from raw input images. They use convolution matrix (kernel) for blurring, sharpening, embossing and edge detection. The lower layers identify general features such as corners and edges, while deep layers extract features specific to the organisms. Example of the CNN is shown in figure \ref{Vgg16}, which shows deep layers of VGG-16 model. With such high extraction power of CNNs, we can use them to extract deep learning features which can be used for detection and classification of SARS-CoV-2. Additionally, the scarcity of SARS-CoV-2 dataset on training CNN can be overcome by the use of transfer learning. Thus, we extract deep learning features from SARS-CoV-2 images using VGG-16 \cite{Vgg16}, Xception \cite{Xception} and DenseNet \cite{DenseNet} which have been pre-trained on ImageNet dataset. The feature maps extracted from different layers are shown in figure \ref{fig:deepFeatures}.

\subsection{Analysis}
The SARS-CoV-2 has strong texture heterogeneity by combining the HOG 
feature and GLCM feature. The shape features and texture features can be 
extracted from the images in SC2-MID which can be combined and used as 
eigenvectors in image classification. The combination of texture and geometric features can describe the surface features of the virus precisely, which contains 
much more abundant information that can help for better classification accuracy 
in image classification.

Moreover, we can leverage strong feature learning power of deep learning networks (CNN) by extracting deep learning features from SARS-CoV-2, which are robust, invariant to translation changes, do not need image pre-processing and can reduce the need of hand crafted features in segmentation, detection and classification of SARS-CoV-2. 
 
\section{Conclusion and Future Work}

The SARS-CoV-2 has some recognizable visual information which can be 
represented by visual features, such as texture and shape features, 
providing a possibility to describe the morphological property of 
SARS-CoV-2 for medical workers.

We may get more electron microscopic images in the 
future to enlarge our dataset. We will establish the evaluation index of the 
SC2-MID. We will extract 
more visual features for image classification in future work. The 
dataset can be used to help medical workers to identify and 
classify the SARS-CoV-2.

\section*{Acknowledgements}

We thank B.E. Jiawei Zhang, due to his great work is considered as important as the 
first author in this paper. We also thank the websites which provide
SARS-CoV-2 images.
%
%
%
%
{\small
\bibliographystyle{splncs04}
\bibliography{Jiawei}
}

\end{document}